\documentclass[usenatbib]{mnras}
\usepackage{graphicx, amssymb, aas_macros, natbib, bm}
\usepackage{ marvosym }
\usepackage{enumitem}
\usepackage{fancybox}

\newcommand{\vmax}{{\it V}_{\rm{max}}}

\newcommand{\mhalo}{{M}_{\rm{halo}}}
\newcommand{\mpeak}{{M}_{\rm{peak}}}
\newcommand{\rvir}{R_{\rm vir}}

\newcommand{\mstar}{{M}_{\star}}

\newcommand{\msun}{{\rm M}_{\odot}}

\newcommand{\lt}{<}
\newcommand{\gt}{>}

\title[Quenched Dwarfs Splish-Splashing in the Field]
{Environmental Quenching of Low-Mass Field Galaxies}

\author[Fillingham et al.]
{Sean P. Fillingham,$^1$\thanks{$\!\!$e-mail: sfilling@uci.edu}\ 
Michael C. Cooper,$^1$\ 
Michael Boylan-Kolchin,$^2$\ 
\newauthor
James S. Bullock,$^1$\ 
Shea Garrison-Kimmel,$^3$\thanks{$\!\!$Einstein fellow}\ 
Coral Wheeler,$^3$\thanks{$\!\!$DuBridge fellow}\ 
\\
$\!\!^1$Center for Cosmology, Department of Physics \& Astronomy,
University of California, Irvine, 4129 Reines Hall, Irvine, CA 92697, USA \\
$\!\!^2$Department of Astronomy, The University of Texas at Austin,
2515 Speedway, Stop C1400, Austin, TX 78712, USA \\
$\!\!^3$TAPIR, Mailcode 350-17, California Institute of Technology,
Pasadena, CA 91125, USA
}

\begin{document}

\pagerange{\pageref{firstpage}--\pageref{lastpage}} 
\pubyear{2018}

\maketitle

\label{firstpage}
\begin{abstract}
In the local Universe, there is a strong division in the star-forming
properties of low-mass galaxies, with star formation largely
ubiquitous amongst the field population while satellite systems are
predominantly quenched. 
This dichotomy implies that environmental processes play the dominant
role in suppressing star formation within this low-mass regime ($
\mstar \sim 10^{5.5-8}~\msun$). 
As shown by observations of the Local Volume, however, there is a
non-negligible population of passive systems in the field, which
challenges our understanding of quenching at low masses. 
By applying the satellite quenching models of Fillingham et al.~(2015)
to subhalo populations in the Exploring the Local Volume In
Simulations (ELVIS) suite, we investigate the role of environmental
processes in quenching star formation within the nearby field.
Using model parameters that reproduce the satellite quenched fraction
in the Local Group, we predict a quenched fraction -- due solely to
environmental effects -- of $\sim 0.52 \pm 0.26$ within $1< R/R_{\rm
  vir} < 2$ of the Milky Way and M31. 
This is in good agreement with current observations of the Local
Volume and suggests that the majority of the passive field systems
observed at these distances are quenched via environmental
mechanisms. 
Beyond $2~\rvir$, however, dwarf galaxy quenching becomes difficult to
explain through an interaction with either the Milky Way or M31, such
that more isolated, field dwarfs may be self-quenched as a result of
star-formation feedback. 

\end{abstract}


\section{Introduction}
\label{sec:intro} 

Recent observations of nearby dwarf galaxies show that low-mass
systems ($\mstar \lesssim 10^{9}~\msun$) currently residing $>1~{\rm
  Mpc}$ from a massive neighbor are almost exclusively star forming
\citep{mateo98, haines08, weisz11a, geha12}.
This is supported by H{\scriptsize I} observations of systems in the Local
Volume, which find a predominately gas-rich field population \citep{spekkens14}.
Together, these results indicate that low-mass systems largely lack
the ability to cease forming stars, or quench, in the field. 
In other words, ``{\it in situ}'' processes, such as morphological
quenching \citep{martig09} or stellar feedback \citep{larson74, ds86}
that operate on more massive field galaxies appear unable to shut down
star formation at the lowest galaxy masses. 

In contrast to the local field population, the low-mass satellites of
the Local Group are nearly universally quenched
\citep[e.g.][]{mateo98, grcevich09, spekkens14}. 
This dramatic difference in the star-forming properties of low-mass
field and satellite galaxies strongly indicates that environmental
mechanisms are responsible for quenching low-mass systems. 
Moreover, the environmental mechanism at play must act with great
efficiency \citep[i.e.~rapidly following infall,][]{fham15, wetzel15b}.
Using $N$-body simulations to model the accretion history of satellites in the
Local Group, \citet{fham16} present a coherent picture of
satellite quenching as a function of satellite and host mass in which
satellites above a host-dependent, critical mass scale are quenched
via starvation while low-mass systems are rapidly quenched via stripping.
This model is supported by complementary observations of satellite populations
in the local Universe, such that it reproduces the fraction of quenched
satellites at $z \sim 0$ over a broad range of masses \citep[e.g.][]{delucia12,
  wetzel13, davies16, stark16}.

Studies of dark matter halo populations within $N$-body simulations,
however, show that a significant fraction of low-mass halos residing
just beyond the virial radius ($R_{\rm vir}$) of a massive halo today
were previously located within $R_{\rm vir}$ \citep{balogh00, mamon04,
  gill05, teyssier12, wetzel14}. 
For example, \citet{GK14} find that these so-called ``backsplash'' halos
comprise roughly $50\%$ of systems in the Local Volume (i.e.~within
$1 < R/R_{\rm vir} < 2$ of the Milky Way).
Given this sizable backsplash population, highly-efficient satellite (or
environmental) quenching, needed to reproduce the Local Group satellite
population at low masses, could be expected to produce a non-negligible
number of quenched galaxies in the field --- potentially in conflict
with current observations. 

In this work, we utilize a suite of $N$-body simulations to investigate the
degree to which environmental quenching models reproduce the observed
population of quenched field galaxies that are currently known to
reside beyond the virial radius of either the Milky Way or M31
(i.e.~in the Local Volume). 
Specifically, does the model of satellite quenching presented by
\citet{fham15, fham16} overproduce the relative number of quenched
systems in the field? 
For more massive galaxies ($\sim10^{9.5}~\msun$), where environmental
quenching is less efficient, \citet{wetzel14} show that observations
agree with the expectations of the model, with $\sim40\%$
of systems within $\sim2~\rvir$ of local groups and clusters likely
quenched by environmental effects.
In Section~\ref{sec:data}, we detail the observational and simulation
data used in this analysis. Additionally we introduce the quenching
models that facilitate comparison between observations and theory.
In Section~\ref{sec:results} and \ref{sec:disc}, we present our
results and discuss any implications and limitations this analysis
might have on the current galaxy evolution paradigm.
Finally, in Section~\ref{sec:endgame}, we summarize this work and
discuss how ongoing efforts will clarify and enhance this framework of
environmental quenching of dwarf satellite galaxies.
Where necessary, we adopt a $\Lambda$CDM cosmology with the following
parameters: $\sigma_{8}=0.801$, $\Omega_{\rm m}=0.266$,
$\Omega_{\rm \Lambda}=0.734$, $n_{\rm s}=0.963$, and
$h=0.71$ \citep[WMAP7,][]{larson11}.

\section{Data and Models}
\label{sec:data}

\subsection{Local Volume Dwarfs}
\label{subsec:obsdata}

Our sample of local dwarf galaxies is drawn from the compilation of
\citet{mcconnachie12}. The dataset builds upon the low-mass satellite sample
from \citet{fham15} by extending beyond the virial radius of the Milky Way and
M31 systems while maintaining the same stellar mass range in order to facilitate
a clean comparison to the known classical satellites of the Local Group.
Our field population is selected to be in the stellar mass range
$\sim 10^{6}-10^{8}~\msun$ and within $1.2~{\rm Mpc}$ of either the Milky Way or
M31.
While this stellar mass range leads to a complete sample of dwarf galaxies
inside the virial radius of the Milky Way and out to $0.5~R_{\rm vir}$ in M31,
the field sample within the Local Volume is potentially incomplete at these
stellar masses.

\citet{whiting07} demonstrate that all-sky surveys using photographic plates are
complete down to a surface brightness of
$\sim 25.5~{\rm mag}~{\rm arcsec}^{-2}$.
This roughly corresponds to And~V ($\mstar \sim 4\times10^{5}~\msun$) at the
distance of M31, suggesting that we are largely complete for stellar masses
above ${\rm a~few}~\times~10^{5}~\msun$ out to distances of at least
$\sim 750~{\rm kpc}$ from the Milky Way.
Inside the footprints of the Sloan Digital Sky Survey \citep{york00} and the
Dark Energy Survey \citep{DES05, DES14}, however, dwarf galaxy samples are
complete well below our lower stellar mass limit of $\mstar = 10^{6}~\msun$ out
to distances of (at least) $1.5~{\rm Mpc}$ from the Milky Way \citep{koposov08,
  tollerud08, walsh09, jethwa16, newton17}.
Finally, while all optical imaging datasets will suffer incompleteness due to
obscuration by the disk of the Milky Way, assuming the dwarf galaxy population
is not biased in a manner where quenched (or star-forming) objects
preferentially reside behind the disk, our results should not be strongly
affected by this incompleteness.

\begin{figure*}
 \centering
 \hspace*{-0.3in}
 \includegraphics[width=7.2in]{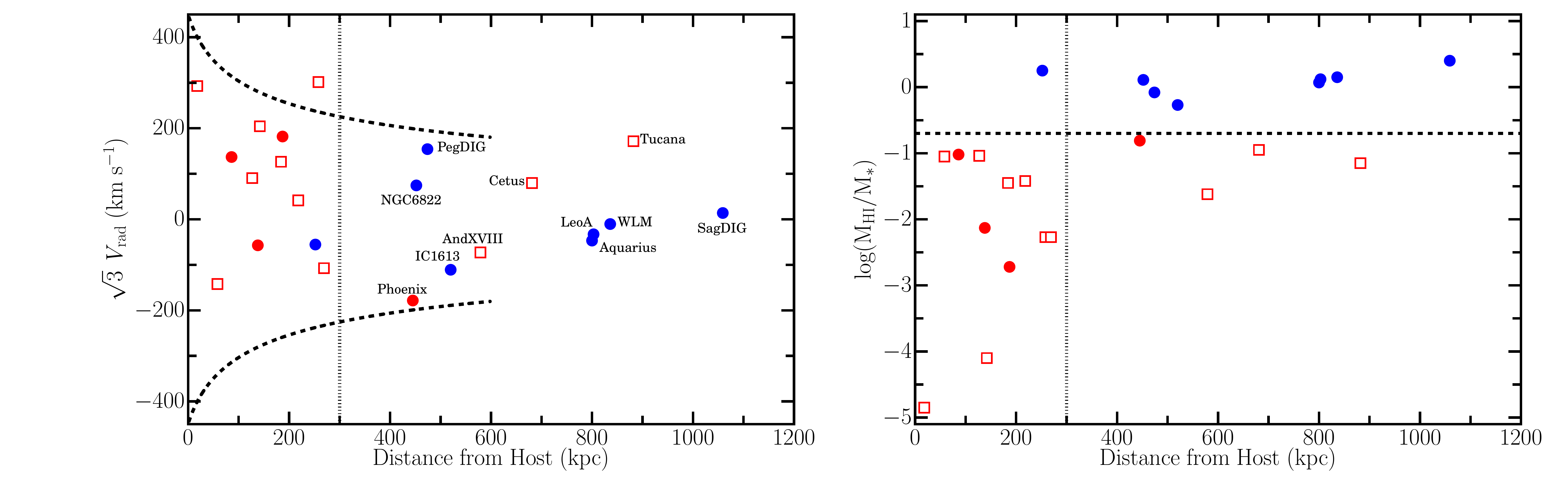}
 \caption{{\it Left}: Host-centric radial velocity (scaled by
   $\sqrt{3}$ to approximately account for tangential motion) for all
   known Local Volume dwarf galaxies in the stellar mass range
   $10^{6}-10^{8}~\msun$ as a function of distance from the nearest
   host (either the Milky Way or M31). The dotted vertical line
   roughly corresponds to the virial radius of a Milky Way-like host
   and the dashed lines correspond to the boundary between bound and
   unbound systems. Satellite points are color-coded according to
   their observed H{\scriptsize I} gas fraction, with blue (and red)
   points denoting gas-rich (and gas-poor) systems, respectively. 
   The closed points signify H{\scriptsize I} detections, while the
   open points correspond to upper limits on the 
   total H{\scriptsize I} mass. {\it Right}: The gas fraction
   ($M_{\rm H\protect\scalebox{0.55}{\rm I}}/\mstar$) as a function of distance
   from the nearest host for the same dwarf galaxies as in the left
   panel. The horizontal dashed line corresponds to a gas fraction of
   $0.2$, below which systems are considered quenched.}
 \label{fig:LVdwarfs}
\end{figure*}

Our final sample includes $11$ dwarf galaxies in the field within
$1.2~{\rm Mpc}$ in addition to the previously-identified $12$
satellite galaxies from \citet{fham15}.
The left-hand panel of Figure~\ref{fig:LVdwarfs} shows the host-centric
radial velocities for our sample \citep{mcconnachie12, tollerud12,
  makarova17}, scaled by $\sqrt{3}$ to approximately account for
tangential motion, as a function of distance to the nearest host
(Milky Way or M31). 
The vertical dotted line at $300~{\rm kpc}$ denotes the approximate location of
the virial radius in a Milky Way-like dark matter halo. Throughout this work, we
adopt $\rvir = 300~{\rm kpc}$ for both the Milky Way and M31.
The dashed lines illustrate the region in which a subhalo is likely
bound to the host, assuming a Navarro-Frenk-White (NFW) dark matter
halo profile \citep{nfw97} with a virial mass of $2\times
10^{12}~\msun$ and a concentration of $8$. 
%


In the right-hand panel of Figure~\ref{fig:LVdwarfs}, we show the H{\scriptsize
  I} gas fraction for each dwarf in our sample as a function of host-centric
distance. The filled points correspond to H{\scriptsize I} detections while the
open points denote upper limits \citep{hunter12, spekkens14, mcconnachie12}.
To separate gas-rich, star forming systems from gas-poor, quenched systems, we
divide the sample based on the H{\scriptsize I} gas fraction
($M_{\rm H\protect\scalebox{0.55}{\rm I}}/\mstar$). Galaxies with an
H{\scriptsize I} gas fraction above $0.2$ are considered star forming, while
galaxies with H{\scriptsize I} gas fractions below $0.2$ are quenched. The
points in Fig.~\ref{fig:LVdwarfs} are color-coded according to this
classification, with blue and red points corresponding to star forming and
quenched dwarfs, respectively.
Our resulting quenched fraction in the local field is
$f_{\rm quench} = 0.36~\pm~0.15$, assuming a binomial error on the measured
quenched fraction.
The measured quenched fraction is largely independent of host-centric distance,
with $f_{\rm quench} = 0.40~\pm~0.22$ within $1 < R/\rvir < 2$ and
$f_{\rm quench} = 0.33~\pm~0.19$ within $2 < R/\rvir < 4$, which suggests that
our sample is not dramatically incomplete for passive systems at large
distances.
While varying the classification threshold for quenched versus star-forming
systems will have a mild impact on the measured quenched fraction, it does not
significantly change the qualitative results of this analysis.

\begin{figure*}
 \centering
 \hspace*{-0.1in}
 \includegraphics[width=6in]{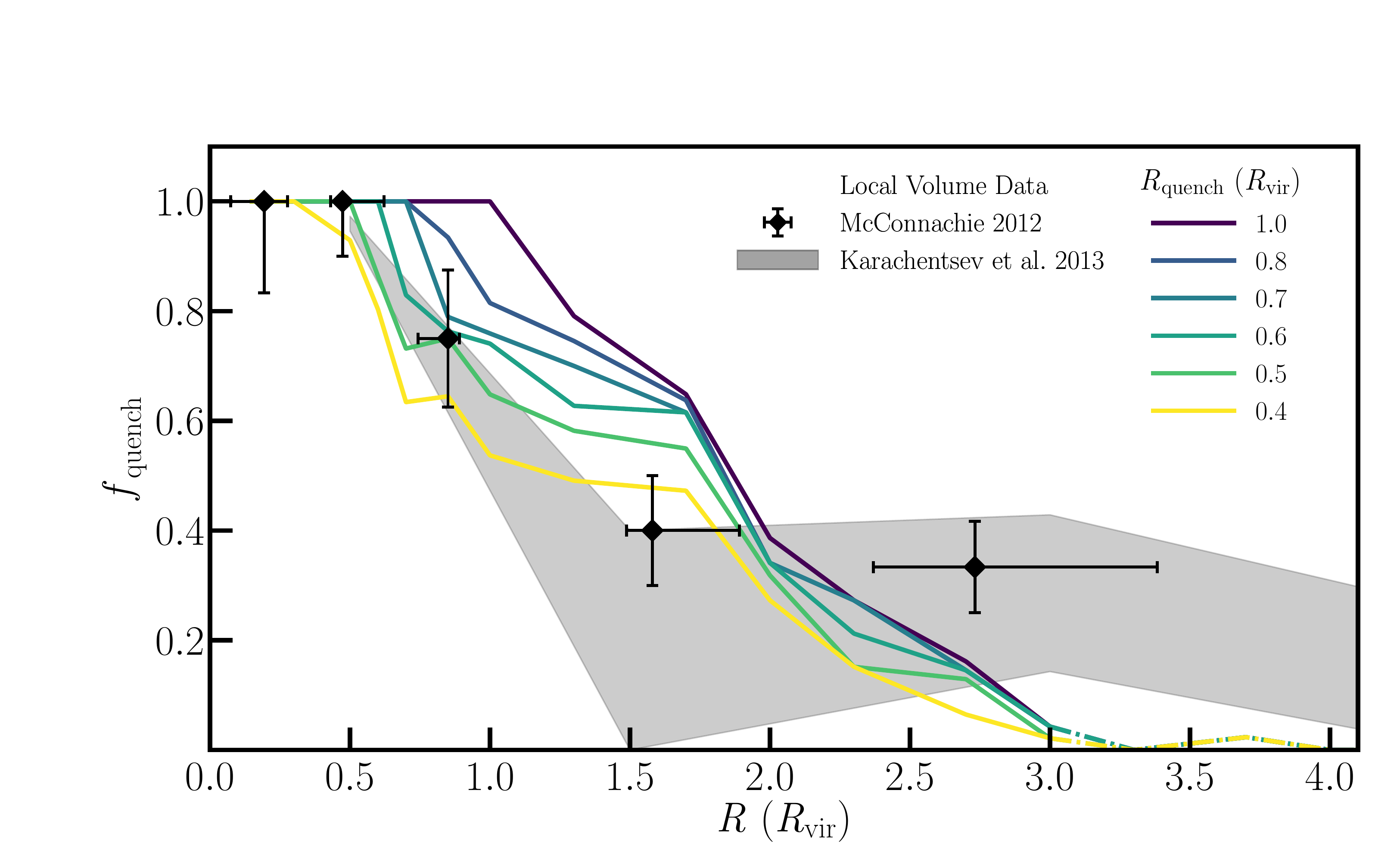}
 \caption{The quenched fraction as a function of host-centric distance for the
   $R_{\rm quench}$ model and Local Volume dwarfs. The black diamonds show the
   quenched fraction for our sample of Local Volume dwarfs (see
   Fig.~\ref{fig:LVdwarfs}) spanning five radial bins, with error bars on the
   host-centric distance corresponding to the $5^{\rm th}$ and $95^{\rm th}$
   percentiles of the distribution in each respective bin and the error bars on
   $f_{\rm quench}$ denoting the $2\sigma$ uncertainty assuming binomial
   statistics. The grey shaded region illustrates a complementary measurement of
   the quenched fraction in the Local Volume determined using the UNGC. The
   upper bound is determined by assuming that both elliptical and transitional
   morphologies are quenched systems, while the lower bound assumes that only
   elliptical systems are quenched. Finally, the colored lines show the inferred
   quenched fraction in the ELVIS suite when varying the radius at which a
   subhalo is considered quenched, $R_{\rm quench}$. Beyond $3~\rvir$
   the model lines are dot-dashed to illustrate the point at which
   some of the simulations in the ELVIS suite are contaminated by low
   resolution particles such that our modeling is less reliable.
   At $R < 2~\rvir$, there is excellent agreement between the
   observed quenched fraction and a model with 
   $R_{\rm quench}=0.5~\rvir$, such that all low-mass dwarfs within
   $\sim2~\rvir$ of the Milky Way and M31 can be explained via environmental
   quenching. Beyond $\sim2~\rvir$, the models cannot explain the
   observed quenched fraction such that these objects are likely
   self-quenching in the field. }
 \label{fig:rq}
\end{figure*}

As an independent measure of the quenched fraction, we also include
data from the Updated Nearby Galaxy Catalog
\citep[UNGC,][]{karachentsev13a}.
Objects are selected according to the same stellar mass limits
($10^{6}-10^{8}~\msun$), given their $K$-band magnitude and assuming a
mass-to-light ratio of $M/L = 1$ \citep{bell01}.
Due to the heterogeneity of star formation rate measurements within the UNGC, we
instead use the morphology of each system as an indicator of its current
star-forming activity. As an upper bound to the quenched fraction, we assume
that both elliptical and transitional morphologies correspond to quenched
galaxies, while the lower bound for $f_{\rm quench}$ assumes only galaxies with
elliptical morphologies are quenched.
For each galaxy in the UNGC, which covers the entire Local Volume, we compute
the distance to M31 and the Milky Way, adopting the lesser of the two as the
host-centric distance.
We then compute the quenched fraction, $f_{\rm quench}$, as a function of
host-centric distance (see the grey shaded regions in Figures~\ref{fig:rq},
\ref{fig:tq}, and \ref{fig:destruct}).
Our measured quenched fraction within the Local Volume is in good
  agreement with the observed fraction of early-type galaxies in the vicinity of
  other nearby massive hosts \citep{karachentsev15c}.

\subsection{Simulations}
\label{subsec:simdata}

The Local Group and its environs provide a unique laboratory for studying the
details of how star formation in low-mass field and satellite galaxies is
affected by the host environment.  In order to take full advantage of this
opportunity, a comparably specific suite of simulations is needed in order to
understand both the details of the host environment and satellite population.
That is, an unbiased comparison of observations in the Local Volume to galaxy
formation and evolution theory requires simulations that span the entire Local
Volume.

The Exploring the Local Volume in Simulations (ELVIS) project is a
suite of cosmological zoom-in dark matter-only simulations and is
comprised of $24$ Milky Way-like hosts as well as $12$ Local Group-like
pairs \citep{GK14}.
Each simulated Local Volume provides complete halo catalogs for
$\mhalo~\gt~2\times 10^{7}~\msun$ and $\vmax~>~8~{\rm km}~{\rm s}^{-1}$ within a
high-resolution region spanning $2-5~{\rm Mpc}$, so as to enable tracking of the
orbital and accretion histories of all dark matter subhalos that could
potentially host a $\mstar > 10^{6}~\msun$ dwarf galaxy. In addition, the ELVIS
suite of simulations is uncontaminated by lower resolution particles out to at
least $3~\rvir$ for each host, such that they reliably trace the properties of
the nearby field dark matter halo population.

From ELVIS, we select dark matter halos and subhalos in the range
$\mpeak = 5\times 10^{9} - 6\times 10^{10}~\msun$, which corresponds to
$\mstar = 10^{6} - 10^{8}~\msun$ via the abundance matching (or stellar
mass-halo mass) relation of \citet{GK14}.
Varying the abundance matching prescription only has a minor affect on the
typical infall times for subhalos in this mass regime and thus a
negligible impact on our results \citep{fham15}. 
In order to increase the precision at which we track subhalo accretion events,
we interpolate, using a cubic spline, all of the dark matter halo
properties in the ELVIS catalog, following \citet{fham15}.
%


\subsection{Quenching Models}
\label{subsec:models}

The physical mechanisms that act upon a galaxy while in the vicinity
of a more massive host are thought to suppress star formation through
modification of the gas reservoir itself or by preventing the gas
reservoir from replenishing its supply. 
However, there are many different potential mechanisms that can
accomplish this and distinguishing which mechanism is operating in
each satellite stellar mass regime can be challenging.
In order to constrain the possible quenching mechanisms, our recent work has
developed subhalo accretion-based models in $N$-body simulations that
should capture the global properties of the quenching process
and constrain the typical quenching timescale \citep{fham15, wheeler14}.
What follows is a brief description of both quenching models used in
this work -- see \citet{fham15} for a more detailed description.

\subsubsection{$R_{\rm quench}$ Model}
\label{subsubsec:rps}

The first model that we examine is based on a ``ram pressure-like'' quenching
scenario where quenching will only occur once a satellite reaches a sufficient
host circumgalactic medium (CGM) density and/or a high enough velocity relative
to the host frame-of-reference, such that the ram pressure experienced by a
satellite is capable of disrupting its gas reservoir.
Both the host CGM density and typical satellite velocity scale with host-centric
distance, such that a model which includes a radial quenching dependence can
broadly capture how these processes should affect the subhalo (i.e.~satellite
galaxy) population.

Within this model, any subhalo that crosses within the quenching radius
($R_{\rm quench}$) is instantaneously and permanently quenched, regardless of
where the subhalo resides today.
In contrast, those halos that never pass within the quenching radius remain
star forming.
By varying the quenching radius ($R_{\rm quench}$), in a manner similar to
\citet{fham15}, the model produces different quenched populations that can then
be compared to the observed quenched fractions in the Local Volume, as shown in
Figure~\ref{fig:rq}.
This model naturally leads to a quenched fraction of unity inside of the adopted
quenching radius.

\subsubsection{$\tau_{\rm quench}$ Model}
\label{subsubsec:starve}

The $\tau_{\rm quench}$ model assumes that quenching occurs at some time,
$\tau_{\rm quench}$, after a dark matter halo crosses within the virial radius of
the host (i.e.~once it becomes a subhalo).
In other words, the adopted quenching timescale ($\tau_{\rm quench}$) sets how
long a dark matter halo must remain a subhalo before it is considered quenched.
This model is able to approximate a ``starvation-like'' scenario,
where the infalling satellite galaxy has been cut off from cosmic
accretion and therefore will stop forming stars when it runs out of
its current reservoir of fuel (i.e.~H{\scriptsize I}+H$_{2}$ gas). 
For relatively short quenching timescales, however, the model may also
mimic suppression of star formation via ram-pressure stripping
(e.g.~where $\tau_{\rm quench}$ roughly follows the crossing time of
the host system).

\begin{figure*}
 \centering
 \hspace*{-0.1in}
 \includegraphics[width=6in]{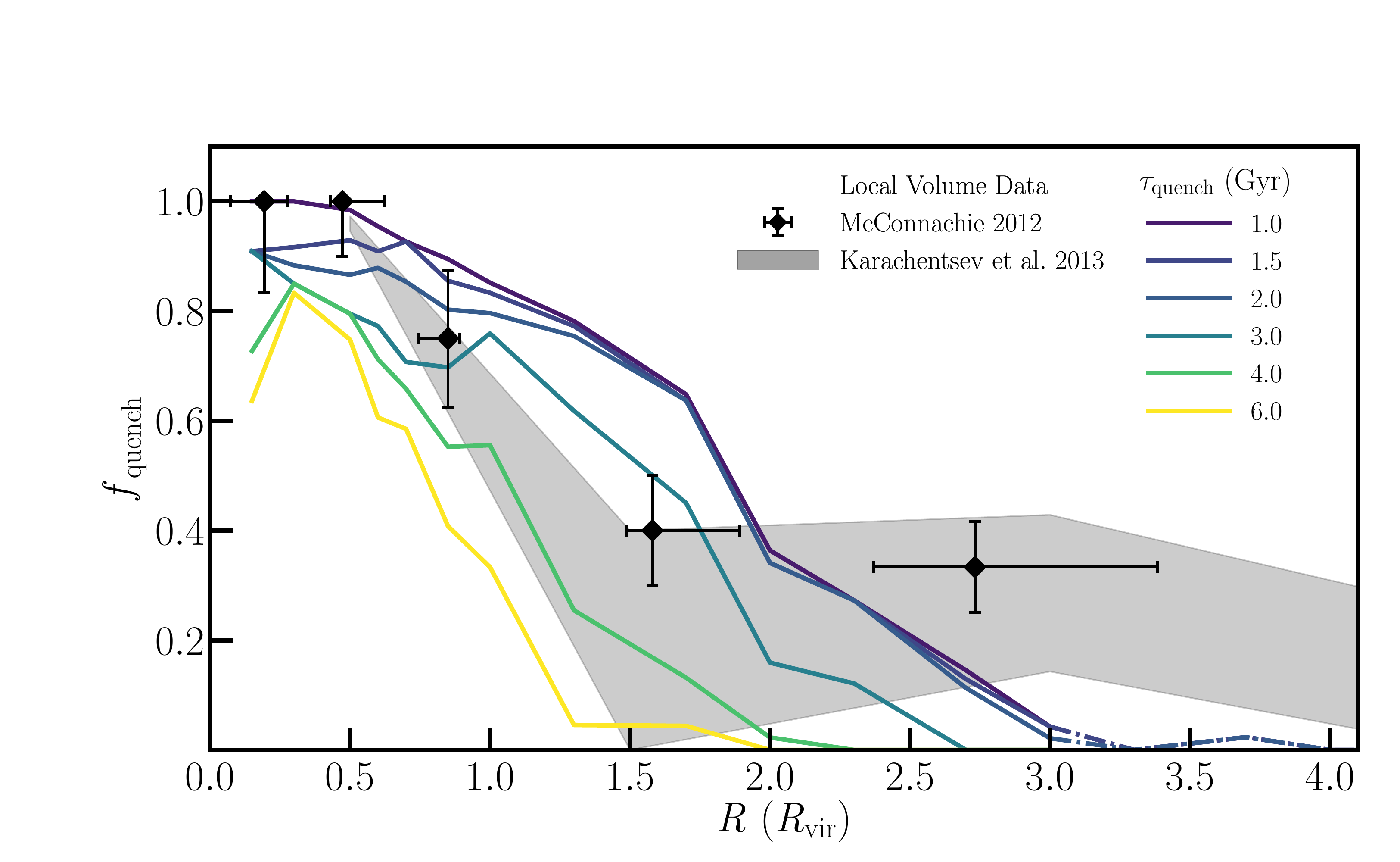}
 \caption{The quenched fraction as a function of host-centric distance for the
   $\tau_{\rm quench}$ model and Local Volume dwarfs.  The black diamonds show
   the quenched fraction for our sample of Local Volume dwarfs (see
   Fig.~\ref{fig:LVdwarfs}) spanning five radial bins, with error bars on the
   host-centric distance corresponding to the $5^{\rm th}$ and $95^{\rm th}$
   percentiles of the distribution in each respective bin and the error bars on
   $f_{\rm quench}$ denoting the $2\sigma$ uncertainty assuming binomial
   statistics. The grey shaded region illustrates a complementary measurement of
   the quenched fraction in the Local Volume determined using the UNGC, as
   detailed in Fig.~\ref{fig:rq}. Finally, the colored lines show the inferred
   quenched fraction in the ELVIS suite when varying the quenching timescale
   from $\tau_{\rm quench} = 1-6~{\rm Gyr}$.
   Beyond $3~\rvir$ the model lines are dot-dashed to illustrate the
   point at which some of the simulations in the ELVIS suite are
   contaminated by low resolution particles such that our modeling is
   less reliable. 
   While the satellite quenched fraction in the Local Group favors
   $\tau_{\rm quench}=1.5~{\rm Gyr}$, such a short quenching timescale
   overproduces the observed quenched fraction at $1 \lt R/\rvir \lt
   2$ due to the contribution from backsplashing systems.
   Similar to the $R_{\rm quench}$ model, beyond $\sim2~\rvir$ the
   models cannot explain the observed quenched fraction such that
   these objects are likely self-quenching in the field.}
 \label{fig:tq}
\end{figure*}

Within this model, all subhalos that remain inside the host dark matter halo at
least as long as the quenching timescale are instantaneously and permanently
quenched.
Subhalos that enter and subsequently exit beyond the virial radius in less time
than the quenching timescale remain star forming.
Such halos are rare for short quenching timescales ($<2~{\rm Gyr}$),
however, comprising $<8\%$ of subhalos at $1<R/\rvir<2$ within our
mass range. 
As shown in Figure~\ref{fig:tq}, by varying the quenching timescale, the
$\tau_{\rm quench}$ model makes distinct predictions for how the
quenched fraction depends on host-centric distance. Overall, a shorter
timescale produces more quenched systems.

\section{Results}
\label{sec:results}

For a wide range of quenching radii ($R_{\rm quench}$) and timescales
($\tau_{\rm quench}$), we apply the models described in
\S\ref{subsec:simdata} to the ELVIS subhalo populations. 
For each implementation of a given model, we measure the dwarf
quenched fraction as a function of distance from the nearest host,
extending to at least $3~\rvir$ (or $\gtrsim 0.9~{\rm Mpc}$ for these
Milky Way-like systems).
Figure~\ref{fig:rq} shows the results for the $R_{\rm quench}$ model,
illustrating the quenched fraction, $f_{\rm quench}$, as a function of distance
to the nearest host for values of $R_{\rm quench}$ ranging from $0.4~\rvir$ to
$1~\rvir$.
For comparison, the black diamonds show the observed quenched fraction in the
Local Volume covering $5$ bins in host-centric distance and the grey
shaded region shows the UNGC quenched fraction as described in
\S\ref{subsec:obsdata}. 

Within the satellite population ($R < \rvir$), the observed quenched fraction is
remarkably high ($f_{\rm quench} \sim 0.9$), so as to favor a quenching radius
of $0.5~\pm~0.1~\rvir$ \citep{fham15}.
For a quenching radius on this scale, our $R_{\rm quench}$ model predicts a
field quenched fraction of $f_{\rm quench}=0.52~\pm~0.26$ at $1 < R/\rvir < 2$
(see Fig.~\ref{fig:rq}).\footnote{The reported quenched fraction for both models
  is the mean quenched fraction in that distance range and the uncertainty is
  the $1\sigma$ scatter in the quenched fraction as measured individually for
  each ELVIS host.}
Overall, at $R < 2~\rvir$, our quenching model with $R_{\rm quench} = 0.5~\rvir$
is consistent with the observed field quenched fraction in the Local Volume,
such that all passive systems at these distances can be explained through
interactions with either the Milky Way or M31.
Beyond $2~\rvir$, however, the backsplash fraction decreases dramatically such
that the model predicts a quenched fraction $\lesssim~0.25$ for all quenching
radii.
At these distances from a massive host, the observed quenched fraction begins to
exceed what the models predict by roughly a factor of two. The implications of
this excess are discussed further in Section~\ref{sec:disc}.

For the $\tau_{\rm quench}$ model, the quenched fraction of the Local Group
satellite population (i.e.~at $R < 1~\rvir$) is best reproduced by a quenching
timescale of $\tau_{\rm quench} \sim 1.5~\pm~1~{\rm Gyr}$ \citep{fham15}. As
shown in Figure~\ref{fig:tq}, this relatively short quenching timescale
following infall yields a field quenched fraction of
$f_{\rm quench}=0.65~\pm~0.24$ within $1 < R/\rvir < 2$, due to the large
population of backsplash halos at that host-centric distance.
Compared to observations of dwarf systems at similar distances in the Local
Volume, this slightly overpredicts the fraction of quenched field galaxies.
Meanwhile, at $R > 2~\rvir$, the $\tau_{\rm quench}$ models underpredict the
fraction of passive systems found locally.

Ultimately, both the $R_{\rm quench}$ and $\tau_{\rm quench}$ models do not
dramatically overpredict the field quenched fraction within the Local
Volume. Instead, the models offer a potential explanation for a substantial
number of the quenched dwarf galaxies that are observed outside of the virial
radius of either the Milky Way or M31 today. Both models, however, do struggle
to explain quenched systems beyond $\sim600~{\rm kpc}$ from either of the Local
Group hosts.
%

\section{Discussion}
\label{sec:disc}

Recent work has shown that the vast majority of low-mass dwarf galaxies that
reside in the field are star forming while a significant fraction that reside
near a massive neighbor are quenched \citep[e.g.][]{geha12}.
A likely consequence of this scenario is that all quenched dwarf galaxies
observed in this mass regime ($\mstar\lesssim10^{9}~\msun$) should have their
star formation shut down via external, environmental processes.
There have been many recent studies that test this picture under the assumption
that the low-mass dwarf galaxy population in the Local Group is representative
of the Universe at large \citep[e.g.][]{wheeler14, slater14, weisz15, wetzel15b,
  fham15, fham16}.

However, the environmental-only quenching hypothesis for low-mass galaxies can
be called into question in the Local Volume since there have been numerous
quenched dwarf galaxies discovered beyond the virial radius of their most
massive nearby neighbor.
This has lead some studies to the conclusion that in-situ quenching of field
dwarf galaxies is potentially occurring \citep{makarova17}.
In this work, we have shown that our models of environmental quenching
(i.e. quenching that only occurs inside $R_{\rm vir}$) can lead to a quenched
fraction of $\sim0.5$ beyond the virial radius but within $2~\rvir$.
Beyond $2~\rvir$, it becomes increasingly difficult to explain the
number of quenched galaxies through an interaction with either the
Milky Way or M31.
%


\begin{figure*}
 \centering
 \hspace*{-0.1in}
 \includegraphics[width=6in]{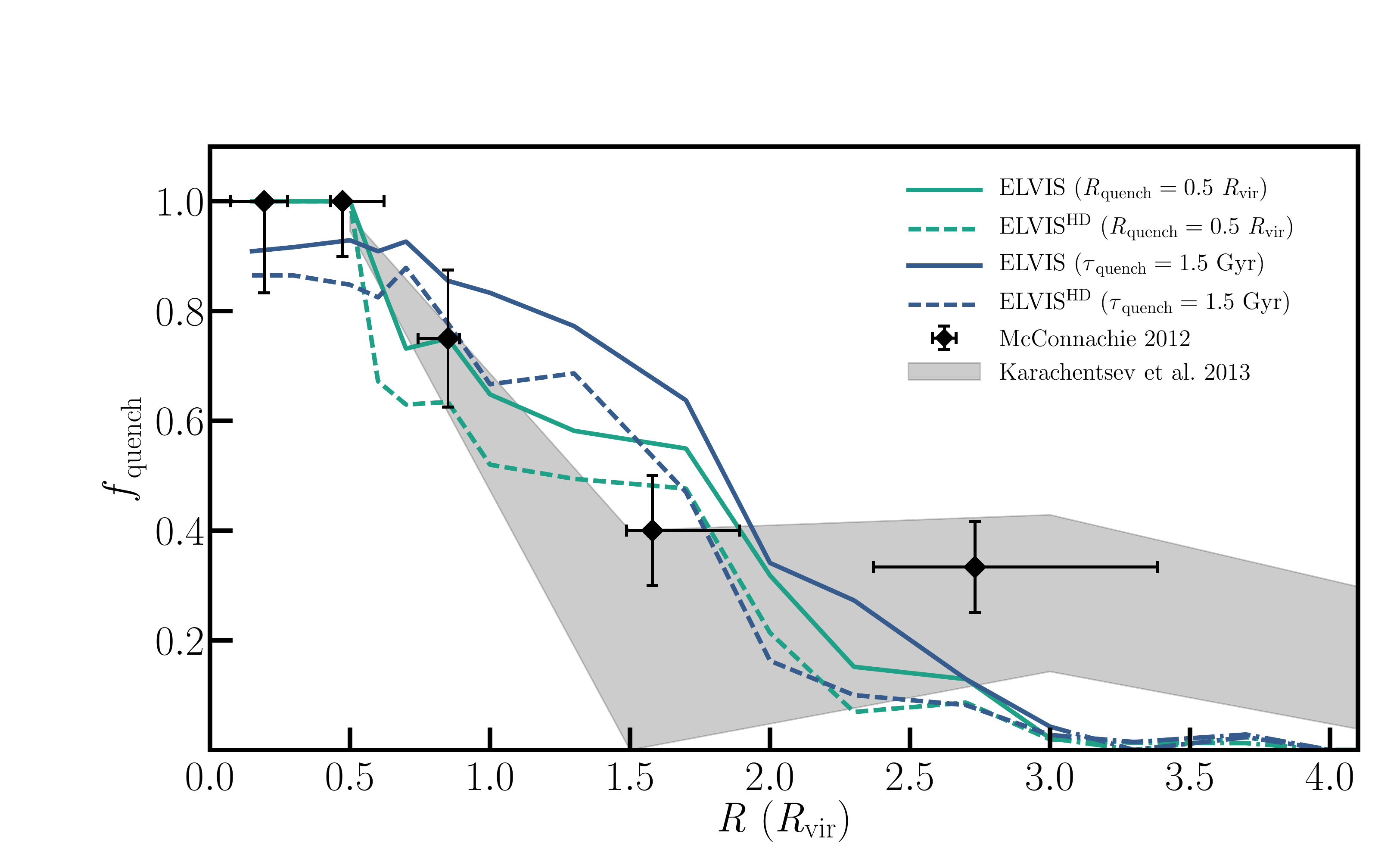}
 \caption{The quenched fraction as a function of host-centric distance for both
   the $\tau_{\rm quench}$ and $R_{\rm quench}$ models in comparison to the
   corresponding measurement in the Local Volume. The solid lines are the result
   of applying the quenching models to the ELVIS dark matter-only simulations,
   while the dashed lines show how our results change when we include the
   effects of tidal disruption by the host potential (ELVIS$^{\rm HD}$, see
   \S\ref{sec:disc}). The black diamonds and grey shaded region illustrate the
   measurements of the quenched fraction in the Local Volume as given in
   Fig.~\ref{fig:rq}.
   As before, beyond $3~\rvir$ the model lines are dot-dashed to illustrate the
   point at which some of the simulations in the ELVIS suite are
   contaminated by low resolution particles.
   The inclusion of subhalo destruction due to tidal effects
   brings both models into better agreement with current observations, such that
   these models can fully explain the observed distribution of quenched dwarf
   galaxies within $2~\rvir$ of both the Milky Way and M31.
   For all of our modeling, quenched dwarf galaxies that currently
   reside beyond $2~\rvir$ cannot be fully explained via environmental
   quenching in either the Milky Way or M31 systems. Further
   emphasizing that self-quenching via star formation feedback is the
   likely quenching scenario in these objects. }
 \label{fig:destruct}
\end{figure*}


\subsection{Baryonic Effects on the Dark Matter Distribution}
\label{subsec:DES}

As shown by \citet{donghia10}, and more recently by \citet{gk17} and
\citet{sawala17}, the baryonic component of the host system can substantially
alter the final subhalo distribution inside the virial radius at $z=0$ (relative
to that found in a pure $N$-body simulation such as ELVIS).
Due to tidal forces, subhalo destruction preferentially occurs in objects with
early infall times and/or more radial orbits.
As such, the distribution of subhalo infall times for a dark matter-only
simulation (such as ELVIS) will be biased towards earlier cosmic times.
Given our quenching models, which are directly connected to the accretion
and orbital history of subhalos, this bias will skew the inferred quenched
fractions as a function of host-centric distance.

In the $R_{\rm quench}$ model, the inclusion of the host system's baryonic
potential will destroy many subhalos that plunge deep into the central regions
of the host halo on highly radial orbits. Given that the majority of these
systems will have pericentric passages within our adopted quenching radius, our
model as applied to ELVIS is likely biased towards greater $f_{\rm quench}$ (see
Fig.~\ref{fig:rq}).
Specifically, we have counted subhalos as ``quenched'' that would have been
destroyed through interactions with a host baryonic potential, therefore
overpredicting the quenched fraction within the Local Volume.
Similarly, in the $\tau_{\rm quench}$ model as applied to ELVIS, we again
overpredict the environmental quenched fraction, since the typical surviving
subhalo in ELVIS spends more time inside the virial radius of the host. Subhalos
accreted at early cosmic time, which are more likely to be tidally disrupted,
are classified as quenched within our model, thereby over-counting the true
number of quenched subhalos that survive to $z=0$.

To demonstrate how subhalo destruction will affect the results presented in
\S\ref{sec:results}, we implement a correction to the ELVIS dark matter
distributions that will broadly capture the tidal effects of the host.
Figures~5 and A2 in \citet{gk17} show the fraction of subhalos that exist as a
function of pericentric distance in two dark matter-only simulations of Milky
Way-like host halos relative to the corresponding subhalo population in
hydrodynamic simulations of the same hosts \citep[using the FIRE model for star
formation and feedback,][]{wetzel16, hopkins14, hopkins17}.
The ratio of subhalos in the dark matter-only versus hydro simulations roughly
follows
\begin{equation}
  N_{\rm DMO} / N_{\rm FIRE} =  40 \, e^{-22 \, d_{\rm peri}/{\rm kpc}} \; ,
\label{eqn:destruct}
\end{equation}
where $N_ {\rm DMO}$ is the number of subhalos that survive to present day in
the dark matter-only simulation, $N_ {\rm FIRE}$ is the corresponding subhalo
count for the hydrodynamic simulation, and $d_{\rm peri}$ is the host-centric
distance at pericenter in kpc.
This radial dependence for subhalo disruption is supported by a larger number of
dark matter-only simulations of Milky Way-like hosts run with an evolving disk
potential (Kelley et al.~in prep).

To mimic the disruption of subhalos in ELVIS, we adopt $(N_{\rm DMO}/N_{\rm
  FIRE})^{-1}$ as the likelihood that a subhalo survives as a function of
pericentric distance.
Beyond $50~{\rm kpc}$, we assume no subhalo destruction.
Within ELVIS, we then randomly destroy subhalos as a function of their
pericentric distance given this probability of survival.
The results of this exercise can be seen in Figure~\ref{fig:destruct} for the
preferred values adopted in both the $R_{\rm quench}$ and $\tau_{\rm quench}$
models. For both models, we find a decrease in the quenched fraction relative to
the same model applied to the dark matter-only simulations, with the difference
being particularly strong for measurements of $f_{\rm quench}$ outside of
$\rvir$.
Overall, the inclusion of tidal effects brings our model predictions into better
agreement with the star-forming properties of dwarfs at $R < 2~\rvir$. 
At $R \gtrsim 2~\rvir$, however, tidal disruption has relatively little impact on the
number of low-mass halos, with environmental processes unable to explain the
suppression of star formation in the most distant dwarfs. 
Finally, it should be noted that the subhalo survival fraction given in
Equation~\ref{eqn:destruct} is derived from a sample dominated by low-mass
systems, such that our subhalo destruction model will likely overpredict the
effect for the ``classical'' dwarfs. A conservative interpretation would be to
treat the results of our tidal disruption models (as shown in
Fig.~\ref{fig:destruct}) as a lower limit on the predicted quenched fraction in
the Local Volume.

\subsection{Quenched Dwarfs in the Field}
\label{subsec:weirdos}

As shown in \S\ref{sec:results}, several of the passive dwarf systems in the
local field population are potentially the product of environmental quenching
mechanisms. 
Specifically, And XVIII, Phoenix, and Cetus currently reside within
$2~\rvir$ of their nearest host, such that they can potentially be
explained with this model. 
\citet{weisz14a} measured the star formation histories (SFH) inferred
from color-magnitude diagrams of resolved stars in Cetus and Phoenix
\citep[see also][]{hidalgo,monelli10a,monelli10b,monelli12}. 
Both objects are consistent with a quenching event occurring roughly
$2~{\rm Gyr}$ ago. 
Additionally, \citet{makarova17} inferred a relatively recent star
formation event in And XVIII, approximately $1.5~{\rm Gyr}$ ago.
The quenched backsplash subhalos in our models that are currently located at
similar distances from their most massive neighbor have a mean lookback time to
infall of $t_{\rm infall} = 5.5 \pm 1.5~{\rm Gyr}$. Assuming a quenching
timescale ($\tau_{\rm quench}$) of $\sim 2~{\rm Gyr}$, this suggests a quenching
time of $\sim 2-5~{\rm Gyr}$ ago, consistent with the quenching times inferred
from the measured SFHs of Cetus, Phoenix, and And XVIII.
Lower-mass quenched systems within $2~\rvir$ \citep[e.g.~Eri~II,][]{DES15,
  koposov15} could have been influenced by environmental processes, however it is
likely that reionization is at play in quenching the lowest-mass dwarfs
independent of environment (\citealt{brown14}; Wimberly et al.~in prep).

At $>2~\rvir$, there are several passive dwarfs with
$\mstar \sim 10^{6}-10^{8}~\msun$ in the Local Volume: Tucana, KK258, KKs3, and
KKR 25 \citep{lavery92, makarov12, weisz14a, karachentsev14b, karachentsev15b,
  karachentsev15a}. These are very unlikely to have been quenched as a result of
interaction with either the MW or M31. Instead, star formation in these systems
was most likely suppressed via in-situ processes. One possibility is
highly-efficient feedback, relative to the depth of the dark matter potential,
that sufficiently disrupts the gas reservoir such that star formation is halted
(if not completely shut down). 
 Dwarf galaxies quenching in isolation (i.e.~in the field),
 independent of environment, have been found in hydrodynamic
 simulations and can offer insight into the mechanisms responsible for
 shutting down star formation in these isolated dwarf galaxies
 \citep[e.g.][]{fitts17}.

\section{Summary}
\label{sec:endgame}

Based on our analysis of quenching models applied to high-resolution $N$-body
simulations, we show that the highly-efficient environmental quenching, required
to suppress star formation in the Local Group satellite population, does not
overpredict the fraction of quenched systems in the local field. Within
$2~\rvir$ of the Milky Way and M31, the passive population at
$10^{6} < \mstar/\msun < 10^{8}$ is consistent with being quenched by
environmental processes.
Moreover, beyond the Local Volume, we conclude that environmental mechanisms are
likely at play in quenching a significant fraction of the low-mass
($\mstar \lesssim 10^{8}~\msun$), passive field systems observed in the local
Universe \citep[see also][]{simpson18}.
Passive dwarfs at $R \gtrsim 2~\rvir$, such as Tucana or KK25, are unlikely to
have been quenched due to interaction with the Milky or M31. Instead, these
systems may represent a tail of the star-forming field population, quenched by
highly efficient feedback \citep[e.g.][]{nihao1, dicintio17, fitts17}.
Altogether, our work adds further support to the model of satellite quenching
outlined by \citet{fham15, fham16}, in which low-mass
($\mstar \lesssim 10^{8}~\msun$) satellites are rapidly quenched following
infall onto a Milky Way-like host.

In conclusion, it should be noted that the broader impact of our work relies on
the assumption that the star-forming properties of the Local Group and
surrounding Local Volume are cosmologically representative \citep[i.e.~are not
atypical,][]{bk16}. 
Studies of low-mass satellites in other nearby systems, however, find a
similarly high quenched fraction like that observed in the Local Group
\citep[e.g.][]{kaisin13}. 
Moreover, by stacking photometric measurements of satellite
populations surrounding a large sample of local Milky Way-like host
systems in wide-field imaging datasets, Phillips et al.~(in prep)
measure a satellite quenched fraction of $f_{\rm quench} \sim 0.7$ at
$\mstar \sim 10^{6-9}~\msun$ in broad agreement with that observed in
the Local Group. 
On the other hand, observations of the NGC~4258 group find a significant number
of blue, likely star-forming, satellites with $\mstar < 10^{8}~\msun$
\citep{spencer14}. In addition, recent work by \citet{geha17} to
spectroscopically identify a large population of low-mass satellites orbiting
Milky Way analogs finds very few passive systems, suggesting that the high
satellite quenched fraction within the Local Group may be very atypical. Without
question, more work remains to fully place the Milky Way (and more broadly the
Local Group) in a cosmological context.

\section*{acknowledgements} 
We thank Tyler Kelley, Dan Weisz, Josh Simon, Andrew Wetzel, and
Walden Cassotto for helpful discussions regarding this work.
We also thank the anonymous referree for providing comments that
helped clarify our work.
This work was supported in part by NSF grants AST-1518257,
AST-1517226, AST-1009973, and AST-1009999.
Additional support was also provided by NASA through grants AR-12836,
AR-13242, AR-13888, AR-13896, GO-14191, and AR-14289 from the Space Telescope
Science Institute, which is operated by the Association of Universities for
Research in Astronomy, Inc., under NASA contract NAS 5-26555.
MBK acknowledges support NASA through grant
NNX17AG29G.
Support for SGK was provided by NASA through Einstein Postdoctoral
Fellowship grant number PF5-160136 awarded by the Chandra X-ray
Center, which is operated by the Smithsonian Astrophysical Observatory
for NASA under contract NAS8-03060.
CW was supported by the Lee A. DuBridge Postdoctoral Scholarship in
Astrophysics. 

This research made extensive use of {\texttt{Astropy}},
a community-developed core Python package for Astronomy
\citep{astropy13}.
Additionally, the Python packages {\texttt{NumPy}} \citep{numpy},
{\texttt{iPython}} \citep{ipython}, {\texttt{SciPy}} \citep{scipy}, and
{\texttt{matplotlib}} \citep{matplotlib} were utilized for our data
analysis and presentation.

\bibliography{splash}

\label{lastpage}
\end{document}